\documentclass[preprint,aps,,prx]{revtex4-1}
\usepackage{ifpdf}
\usepackage{times}
\usepackage{amsmath}
\usepackage{amssymb}
\usepackage{units}
\usepackage{stmaryrd} 
\usepackage[T1]{fontenc}
\usepackage[applemac]{inputenc} 	
\usepackage{graphicx}
\usepackage{dcolumn}
\usepackage{bm}
\usepackage{color}

\def\be{\begin{equation}}
\def\ee{\end{equation}}
\def\bea{\begin{eqnarray}}
\def\eea{\end{eqnarray}}

\newcommand{\RTO}{$\mbox{Rb}_{2}\mbox{Ti}_2\mbox{O}_5$}
\newcommand{\RTOd}{$\mbox{Rb}_{2}\mbox{Ti}_2\mbox{O}_{5-\delta}$}

\newcommand{\MTO}{$\mbox{M}_{2}\mbox{Ti}_2\mbox{O}_5$}

\newcommand{\KTOd}{$\mbox{K}_{2}\mbox{Ti}_2\mbox{O}_{5-\delta}$}

\begin{document}
\preprint{0}

\title{$\mbox{Rb}_{2}\mbox{Ti}_2\mbox{O}_{5-\delta}$: A superionic conductor with colossal dielectric constant}

\

\author{Rémi Federicci\textsuperscript{1},   Stéphane Holé\textsuperscript{1}, Aurelian Florin Popa\textsuperscript{2}, Luc Brohan\textsuperscript{2},   Benoît Baptiste\textsuperscript{3}, Alexis S. Borowiak\textsuperscript{4}, Silvana Mercone\textsuperscript{4,1} and Brigitte Leridon\textsuperscript{1}*} 
\affiliation{\textsuperscript{1}\textit{LPEM-ESPCI Paris, PSL Research University, CNRS, Sorbonne Universités, UPMC, 10 rue Vauquelin, F-75005 Paris, France}} 
\affiliation{\textsuperscript{2}\textit{Institut des Matériaux Jean Rouxel (IMN), Université de Nantes, CNRS, 2 rue de la Houssinière, BP 32229, F- 44322 Nantes Cedex 3, France}} 
\affiliation{\textsuperscript{3}\textit{IMPMC,  Sorbonne Universités UPMC, CNRS, 4 place Jussieu, F-75005 Paris, France}}
\affiliation{\textsuperscript{4}\textit{Université Paris 13, Sorbonne Paris Cité, LSPM,  CNRS, 99 Avenue J.-B. Clément, F- 93430 Villetaneuse, France}}
\affiliation{*\textbf{Corresponding author:} brigitte.leridon@espci.fr}

\begin{abstract}
Electrical conductivity and high dielectric constant are in principle self-excluding, which makes the terms insulator and dielectric usually synonymous. 
This is certainly true when the electrical carriers are electrons, but not necessarily in a material where ions are extremely mobile, electronic conduction is negligible and the charge transfer at the interface is immaterial.  
Here we demonstrate in a perovskite-derived structure containing five-coordinated Ti atoms, a colossal dielectric constant (up to $\mbox{10}^9$) together with very high ionic conduction $\mbox{10}^{-3}\mbox{S.cm}^{-1}$ at room temperature. Coupled investigations of  I-V and dielectric constant behavior allow to demonstrate that, due to ion migration and accumulation, this material behaves like a giant dipole, exhibiting colossal electrical polarization (of the order of $\mbox{0.1\,C.cm}^{-2}$).  Therefore, it may be considered as a "ferro-ionet" and is extremely promising in terms of applications. 

\end{abstract}

\maketitle

\textbf{
The perovskite structure was initially discovered in CaTiO$_3$  by Gustav Rose in 1839 and named after the famous mineralogist Lev Perovski.  Ever since then, perovskites - as well as their derived structures - have demonstrated outstanding properties. These span for instance from ferroelectricity and piezoelectricity  in $\mbox{BaTiO}_3$  and in Pb(Zr$_x$Ti$_{1-x})$O$_3$  \cite{VonHippel:1946, Wul:1946, Jaffe:1971,Cohen:1998},  to unrivaled high critical temperature for superconductivity in the cuprates \cite{Bednorz:1986,Chu:1987}, and include also colossal magnetoresistance in the manganites \cite{Ramirez:1997}. 
Among the derived structures from perovskites, the \MTO\, family (with M an alkaline element) constitutes a peculiar class of materials. As a matter of fact, the titanium atoms in these materials are of 5-coordination, which initially raised the question of their structure and chemical stability in the sixties  \cite{ANDERSSON:1960fi,Andersson:1961vf}, and the structure is lamellar, with planes of pyramids separated by the alkaline elements.
We present here the first electrical study of \RTO\ and demonstrate extremely unusual and interesting properties. In particular we show that, when oxygen vacancies are created in the material,  the dielectric constant and the electrical polarization reach unprecedented values such as $\mbox{10}^9$ and $\mbox{0.1 C.cm}^{-2}$ respectively, which we attribute to high ionic conduction of the order of $\mbox{10}^{-3} \mbox{ S.cm}^{-1}$, together with extremely low electronic conductivity (lower than 10$^{-8}$S.cm$^{-1}$).  These two combined properties make the material simultaneously a super-ionic conductor and an extremely good dielectric, promoting this material as an ideal candidate for applications, especially in supercapacitors.}\\

\begin{figure}[ht]\centering
\includegraphics[width=1.2\linewidth]{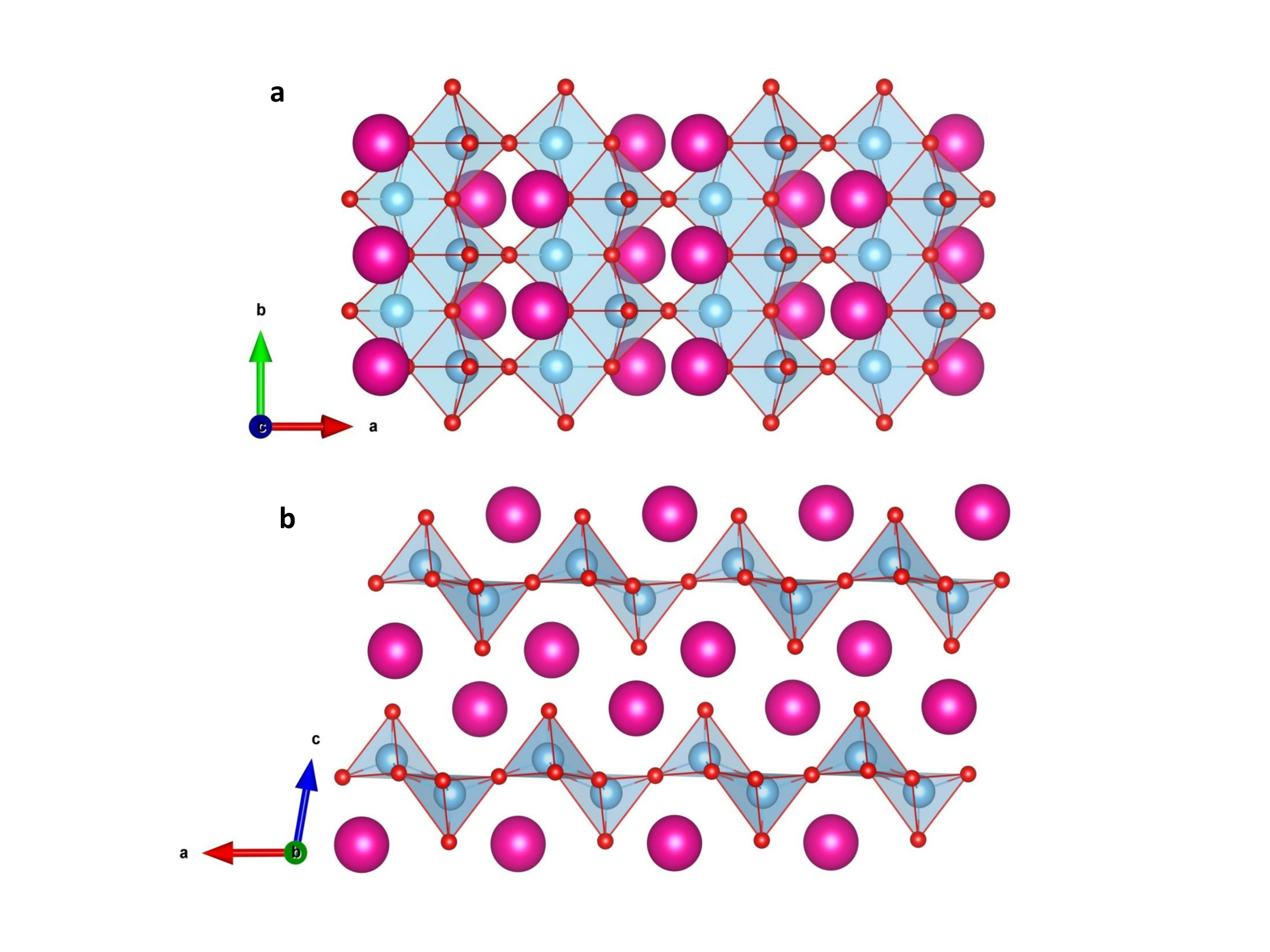}

\caption{Crystal structure of \MTO, visualized using VESTA \cite{VESTA} software. \textbf{(a)}: View along the c-axis. \textbf{(b)}: View along the b-axis, evidencing the lamellar structure. The $(\mbox{Ti}_2\mbox{O}_5)^{2-}$ planes are composed of staggered upside and downside oxygen pyramids that share either a vertex or an edge. The Ti atoms (in blue) are inside the oxygen pyramids (oxygen atoms are in red).  These planes are separated by the M=Rb or K atoms (in pink).}
\label{vesta}
\end{figure}

\begin{figure}[ht]\centering 
\includegraphics[width=1.0\linewidth]{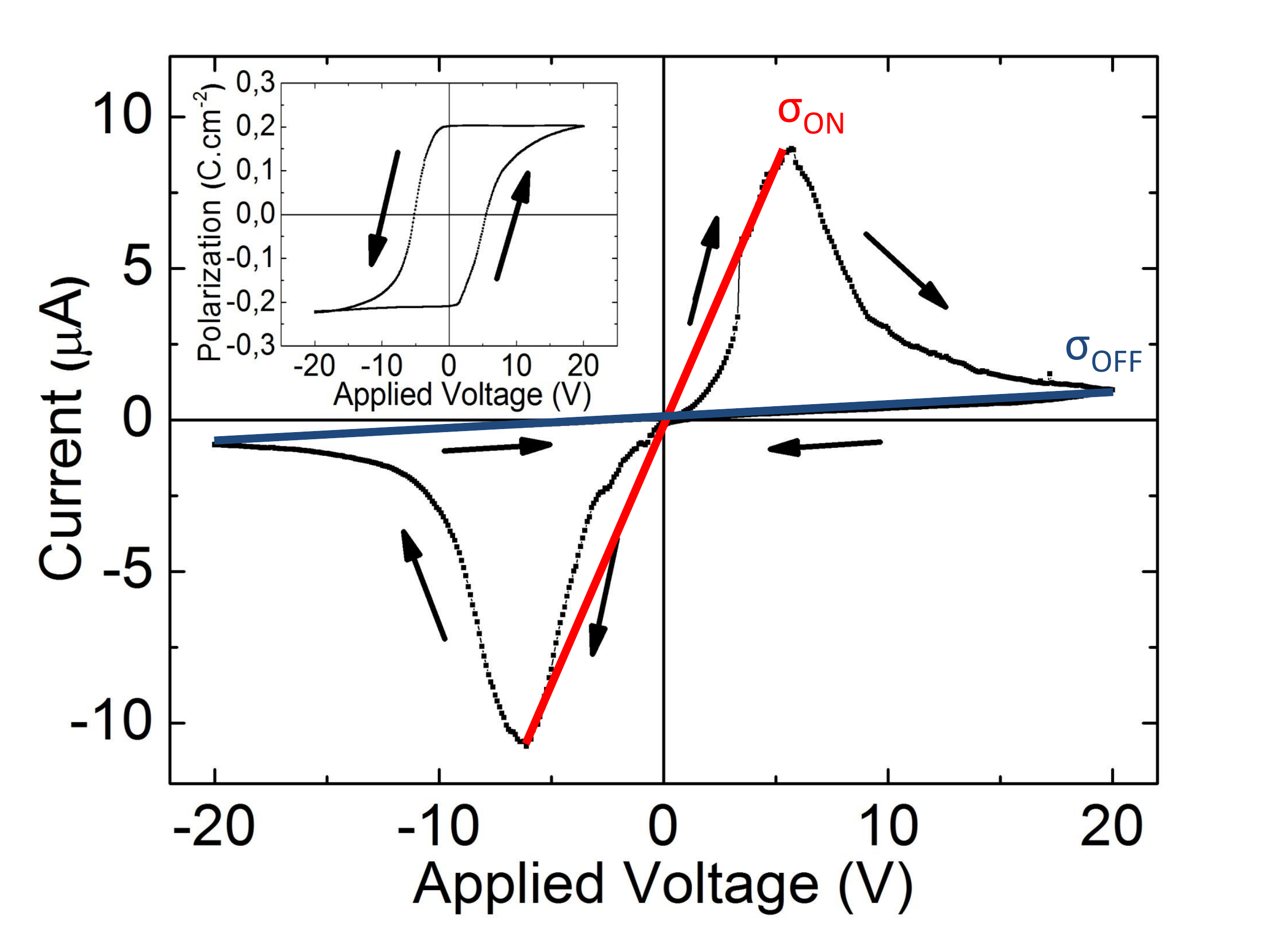}
\caption{Typical measured current versus applied voltage for a single crystal at 260\,K (for a sawtooth voltage with $\omega = 1.2\mbox{mHz}$). For this sample the crystal size is about 200\,$\mu$m $\times$ 400\,$\mu$m $\times$1000\,$ \mu$m.  The peak current density is about 10\,mA.cm$^{-2}$.
Inset: Corresponding polarization versus applied voltage. Note the colossal value of the polarization. For a typical sample, the integrated charge is of the order of 100\,$\mu$C, which  divided by the contact area gives a density of polarization of the order of 0.1\,C.cm$^{-2}$. }
\label{IV}
\end{figure}

\begin{figure*}[ht]\centering 
\includegraphics[width=\linewidth]{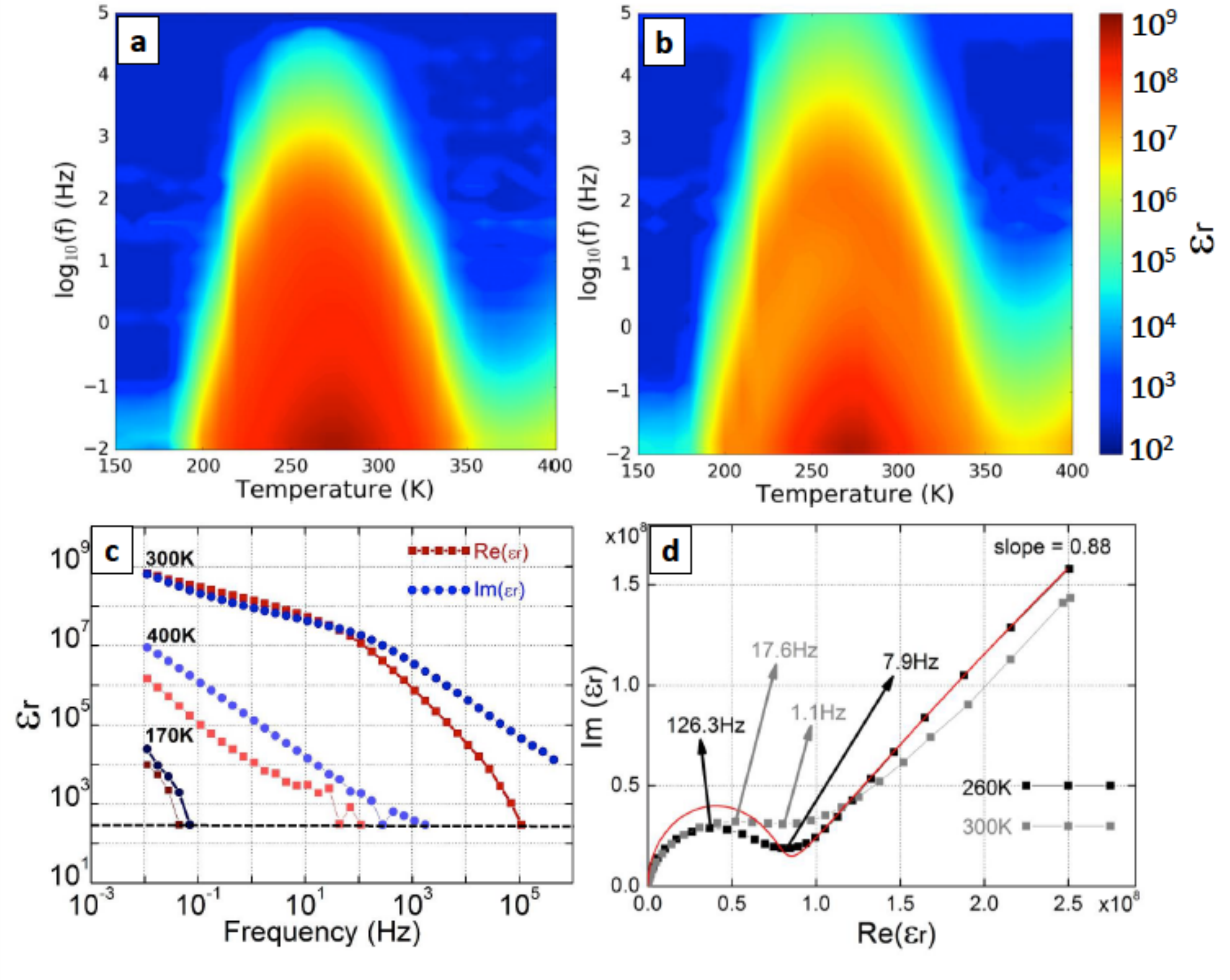}
\caption{\textbf{Top: } Dielectric constant of a helium-annealed monocrystalline sample. \textbf{(a)} Re($\epsilon_r$)  and \textbf{ (b) }Im($\epsilon_r$)  as a function of temperature and frequency with $V_{AC} = 100\,\mbox{mV}$. \textbf{Bottom:  (c)} Re($\epsilon_r$) (red squares) and Im($\epsilon_r$) (blue dots) curves vs frequency for the corresponding temperatures 170\,K, 300\,K and 400\,K. The dotted line represents the measurement threshold.\textbf{ (d)} Cole-cole plot of the permittivity at 300\,K (corresponding to the upper curves in \textbf{(c)}), exhibiting a linear part with a slope of about 0.9. Red line: Electrical model pictured in Fig.\ref{EM}a, with parameters:  R$_I$= 3 $\times$10$^5$\,$\Omega$ ; C$_{max}$=1.5$\times$10$^{-6}$\,F ; C$_w$= 30.7$\times$10$^{-9}$\,F ; W=5$\times$10$^7$\,F.s$^{-1/2}$.  The fact that the slope of the line is close to 1, and that Re($\epsilon_r$) and  Im($\epsilon_r$) tend to $1/\omega^{1/2}$ variation at low frequency points toward a Warburg-type impedance, characteristic of a diffusion process.}.  
   
\label{epsilonr}
\end{figure*}

\textbf{\MTO\, samples}  

Crystals of \MTO\, were synthesized using the technique described in the Supplemental Material.  The crystals take the shape of long transparent needles of typical length about 1\,mm. In Figure \ref{vesta} is displayed the atomic structure of \MTO.  This structure was known to be c2/m for M=K \cite{ANDERSSON:1960fi,Andersson:1961vf}, and we have shown through X-Ray diffraction (XRD) experiments presented in the Supplemental Material that it is also the case for M=Rb. For the electrical measurements, we mostly focussed on \RTO\ samples.


\textbf{Electrical measurements: I-V curves}

Several crystals of \RTO  \,of typical dimensions $200\,\mu \mbox{m} \times 400\,\mu  \mbox{m} \times 1\, \mbox{mm}$ were isolated and contacted using silver wire two-probe contacts glued with carbon paint. (See the photograph in Fig. 1 in the Supplemental Material.)  Typical contact areas were about 0.1\,mm$^2$. Other types of contacts and electrodes were experimented, and all gave similar results (See the corresponding paragraphs in the Supplemental Material).

The electrical resistivity of the as-grown \RTO\ crystals was found to be extremely high (typically above 10$^8\, \Omega$.cm). 
However, after annealing under helium atmosphere at 400\,K for a couple of hours, the resistance at 300\,K went down by at least four orders of magnitude. We hereafter refer to this process as "activation" of the material and will discuss later a possible interpretation.
The material could be subsequently "deactivated" by annealing under oxygen atmosphere , which tends to indicate that the mechanisms at play are redox processes.

In Fig. \ref{IV} is displayed the I-V curve of an "activated" single crystal at 260\,K, for an oscillation frequency of the applied voltage of $1.2\, \mbox{mHz}$. 
This hysteretical curve is typical of what is found in the $200\, \mbox{K}-330\, \mbox{K}$ temperature range, although the current is maximal at 260\,K. Outside this temperature range, by contrast, the I-V curve is typical of a simple small capacitance in parallel with a high resistance corresponding to a resistivity value of about $\mbox{10}^8 - \mbox{10}^9\, \Omega\mbox{.cm}$. (See Fig. 2 of the Supplemental Material.) Please note that the curve is \textit{not self-crossing} and is odd with respect to voltage. The difference between $ \sigma_{ON}$ and $ \sigma_{OFF}$ defined as indicated in Fig. \ref{IV} characterizes the current hysteresis. 

The inset of Fig. \ref{IV} represents the polarization, calculated by integrating over time the measured current and dividing by the area of the contacts, as function of the applied voltage. Such cycles are indeed evocative of a ferroelectric behavior. The colossal value of this polarization, of about 0.2\,C.cm$^{-2}$ is noteworthy. For the whole set of measured samples, the integrated charge was typically of the order of 100\,$\mu$C and the corresponding polarization typically about 0.1\,C.cm$^{-2}$.

\textbf{Relevance of a structural transition}
Since the I-V and P-V curves of Fig. \ref{IV} are strongly evocative of a ferroelectric response, the possible existence of structural transitions at around 330\,K and 200\,K, which could explain the strong polarization of the system, was investigated.  The powder x-ray  diffractograms of \RTO\ at temperatures spanning from 100\,K to 400\,K are displayed in Fig. \,10 of the Supplemental Material, fully consistent with a C2/m centrosymmetric structure and exhibiting no indication for a structural transition.  The obtained lattice parameters are given in Table XI of the Supplemental Material.  Single crystal XRD data are also presented and yield the same results. As a matter of fact, similar results were obtained for a sample activated by vacuum-annealing or N$_2$-annealing at 400\,K  (see Fig. 11 of Supplemental Material).  
These results allowed to discard the possibility of a structural transition toward a conventional ferroelectric state for both as-grown and activated samples at least to the level of precision of the XRD. (XRD data are not able to infer the presence of vacancies at the O sites.)

\textbf{Permittivity measurements}

The dielectric constant as a function of temperature and frequency was also investigated and the results are displayed in  Fig. \ref{epsilonr}.  The complex admittance of the sample was measured using a Solartron 1296 Dielectric interface and a Solartron 1260 impedance/gain phase analyser,  and the corresponding equivalent dielectric constant $\epsilon_r= \epsilon/\epsilon_0$ was extracted.  In Fig. \ref{epsilonr}a the real part of the dielectric constant  is plotted in logarithmic units.  It increases by up to  six orders of magnitude from high to low frequency between 200\,K and 320\,K and reaches a value of more than $10^8$ at 300\,K at 0.01\,Hz. The imaginary part of  $\epsilon_r$, displayed in Fig. \ref{epsilonr}b is found to follow a similar behavior.
In Fig. \ref{epsilonr}c are plotted the real and imaginary parts of $\epsilon_r$ as a function of frequency for three different temperature values. While the curves are featureless at 170\,K and 400\,K, the curves at 300K exhibit two distinct regimes separated by a cut-off frequency $\omega_C$ of about $10\,\mbox{Hz}$. 
 
Remarkably enough,  below this cut-off frequency, the imaginary and real parts of the dielectric constant  seem to converge to the same frequency behavior approximately following $1/\sqrt{\omega}$ (see Fig. \ref{epsilonr}c). This observation is typical of ionic diffusion processes as described by Warburg impedances   \cite{Warburg:1899} evolving as $Z_W \propto (1-j)/\sqrt{\omega}$ in the case of an ideal semi-infinite system with perfectly flat electrode interface.  It is related to the presence of a diffusion layer in the system, whose thickness increases when $\omega$ decreases.   
Therefore this Warburg element points toward an electrolyte nature for the material. The Fig. \ref{epsilonr}d displaying Cole-Cole \cite{Cole:1941fy,Cole:1942bo} plots acquired at 260\,K and 300\,K gives a better picture of the diffusion mechanism.  Below 5\,$-$10\,Hz the linear relation typical of the ionic diffusion is clearly evidenced through the existence of a constant phase element.  

These results were reproduced for many different single crystals and polycrystals of  \RTOd\, (See the supplemental Material) and  were also observed in  \KTOd\, although most of the study was performed on \RTOd.   Here  $\delta$ refers to the reduction process occurring during the annealing.  Its value depends on the annealing conditions and systematic measurements of $\delta$ will be the object of another study. For as-grown samples or samples annealed at 400\,K under oxygen atmosphere, the effects are strongly reduced and the sample is equivalent to a capacitor with a parallel resistance of very high resistivity larger than  $\sim \mbox{10}^{9}\, \Omega\mbox{.cm}$.  (See the corresponding maps for oxygen-annealed crystals in Fig. 4 of the Supplemental Material.) 

In Fig.\,\ref{EM}.a  is displayed another permittivity curve obtained under room atmosphere as function of frequency. In this case, the "Warburg" impedance element is slightly modified and a saturation effect for $\mbox{Re}({\epsilon_r})$ appears at low frequency. The curves indeed exhibits three distinct regimes: the saturation at low frequency, the Warburg regime with slope -1/2 and a conduction regime where $\mbox{Im}({\epsilon_r})$ has a slope -1 with log(f). 

We are then able to propose an equivalent electrical model for the samples, which is pictured in the inset of Fig. \ref{EM}.a. In this equivalent circuit model, $C_{max}$ is the maximal capacitance that can be attained by the sample. This corresponds to the maximum of dielectric constant $\epsilon_r\sim 10^9$ and describes a microscopic situation where all the negative ionic species are stacked at one edge of the crystal and the positive ionic species at the opposite edge. The polarization is then maximal.  This is attained at extremely low frequency and for sufficient value of the electric field , where the crystal behaves like a double-layer super-capacitor. The second capacitance $C_{W}$ in parallel with the Warburg impedance element determines, together with $R_{I}$ the cut-off frequency $\omega_C=1/R_IC_W$ that separates the Warburg regime, where the diffusion layer is experienced by the system, from the regime of freely moving ions characterized by the resistivity $R_{I}$ at high frequency, when edge effects are not relevant any more.  
This electrical model accounts reasonably well for the Cole-Cole plot of Fig. \ref{IV}d.
The conductivity  $\sigma_I=R_{I}^{-1}$  is plotted in Fig. \ref{EM}b together with $\sigma_{ON}$ and $\sigma_{OFF}$ extracted from the I-V curves as function of temperature. The excellent agreement between these values allows to establish without any doubt that the origin of the hysteretic cycles and of the giant polarization is the presence of these extremely mobile ions.

\begin{figure*}[ht]\centering 
\includegraphics[width=1\linewidth]{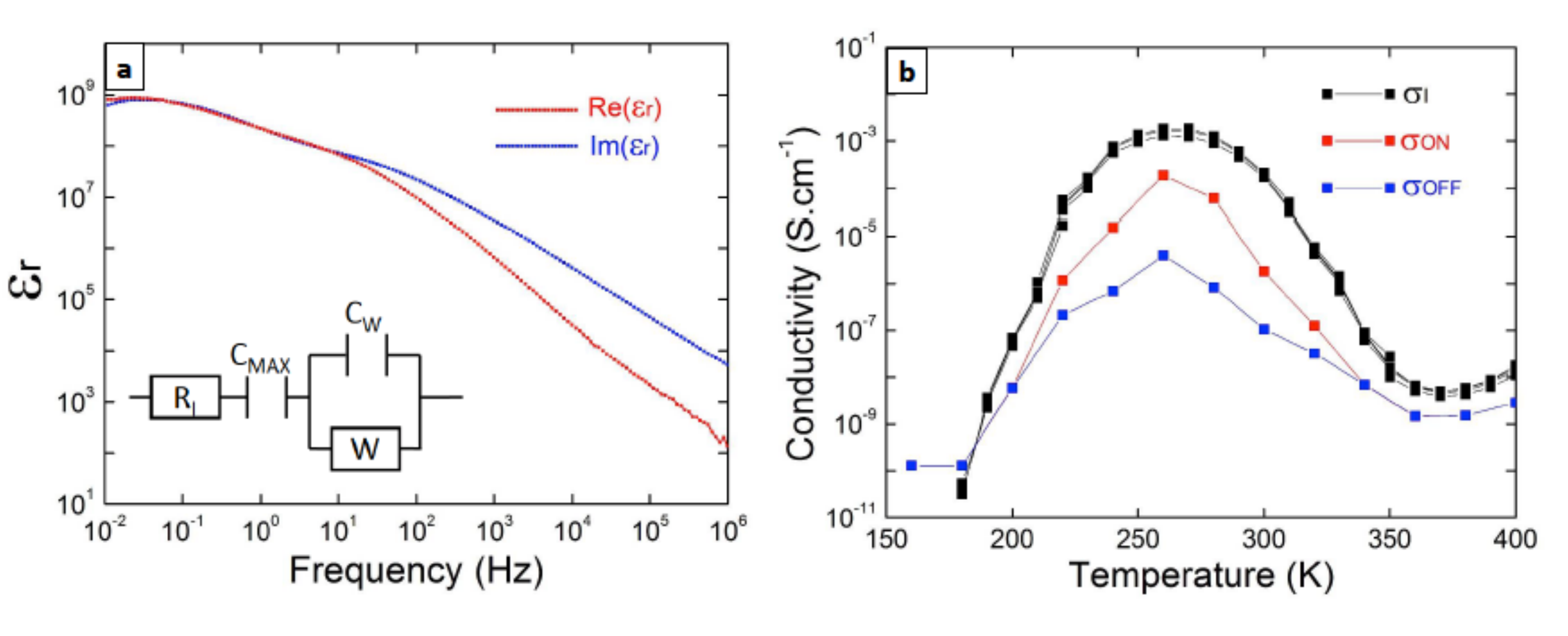}
\caption{ \textbf{(a)}  Real and Imaginary parts of the relative permittivity $\epsilon_r=\epsilon/\epsilon_0$ as function of frequency at 300\,K under room pressure and temperature conditions. Re$(\epsilon_r)$ shows a saturation at low frequency corresponding to accumulation of ionic species on the edges of the sample. Inset: Equivalent electrical model for the samples, accounting for the Cole-Cole plots of Fig. 3d. \textbf{(b)} Ionic conductivity $\sigma_I$ extracted from the high frequency value of Re($\epsilon$) plotted as a function of temperature. $\sigma_{ON}$ and $\sigma_{OFF}$ are extracted from the I-V curves as shown in Fig. \ref{IV}. The difference between  $\sigma_{ON}$ and $\sigma_{OFF}$ characterizes the existence of the ferroelectric-like cycles and the figure evidences the strong correlation between the existence of the hysteretic cycles and the ionic conductivity.}
\label{EM}
\end{figure*}

\textbf{Colossal permittivity due to ionic motion}

The most salient feature of this new material is its colossal dielectric constant (up to $10^9$  at low frequency under room atmosphere). 
This observation is corroborated by independent measurement of the electrical polarization through the I-V data that is also remarkably high and quantitatively consistent with the dielectric constant values. 

This anomalously high value of the polarization suggests the existence of a macroscopic dipole in the sample. 
This dipole is shown to be related to ion displacements in the material as ion diffusion is attested by the presence of a Warburg element as explained above.  It is established that this dipole is maximal between 200\,K and 320\,K, precisely in the range where the ionic conductivity extracted from the equivalent permittivity measurements is maximal. This macroscopic dipole can only exist provided the electrodes are ion-blocking.  This has to be the case, since for any combination of electrodes, the presence of a giant moment and a Warburg element is observed. This Warburg element corresponds to the local accumulation of charges that build up an electric field which corrects the external field. If the electrodes were not ion blocking, such a charge accumulation could not occur. In addition, for evaporated gold electrodes, that are specifically ion-blocking, the same results are obtained. (See Fig.6 in the Supplemental Material).  Another condition for the observation of such a high dielectric constant is that at least two ionic species of opposite signs are involved in the process. If only one were at play, the high capacity would be in series with a low capacity, which would be detrimental to $\epsilon_r$. This may explain indeed why this material is the first ionic conductor where such a high dielectric constant is reported.  

The fact that annealing under oxygen-depleted atmosphere strongly enhances the polarization and the dielectric constant  and annealing under oxygen strongly decreases it (see Fig. 3 in Supplemental Material),  points toward the role of oxygen vacancies.  The presence of O vacancies is also highly suspected since reversible color changes are observed with annealing \cite{Sekiya:2000} (see the last paragraph in the supplemental material.)
Candidates for the migrating ions are therefore the oxygen ions or vacancies but 
further work has to be carried out in order to decipher the exact nature of these ions. As a matter of fact, we cannot discard any possible contamination by water vapor during even brief exposure to atmosphere. This would open the possibility for other mobile ionic species such as protons or hydroxide ions, etc… In any case, the mobility appears to be favored by the presence of oxygen vacancies.  It is also necessary to understand the mechanism for  suppression of motion above 330\,K.

We also  investigated the piezoelectric response of this material in order to confirm the existence of charge displacement and indeed evidenced local piezoelectric response. The results are shown in Fig. 12 of the Supplemental Material.   Such response is indeed expected in any material exhibiting strong spontaneous electrical polarization whatever its microscopic origin. 

Another titanate known for its high dielectric constant is  $\mbox{CaCu}_3\mbox{Ti}_4\mbox{O}_{12}$ (CCTO)  \cite{Ramirez:2000tn,Homes:2001ve}. In this compound, a giant dielectric constant (about $\mbox{10}^{3}-\mbox{10}^{5}$, so still  4 to 5 orders of magnitude smaller) and "relaxor-like" behavior has been evidenced and attributed either to internal barriers  effects \cite{Li:2005ep} or to the presence of oxygen vacancies \cite{Wang:2012ib}. 
However, in these previously discovered giant dielectric constant materials \cite{Ramirez:2000tn,Homes:2001ve,Wu:2002ko, Larsen:1973vd}, the dielectric constant shows a conventional behavior without ionic diffusion processes such as Warburg elements. 

The ionic conductivity $\sigma_I$  extracted from the I-V measurements  is of the order of  10$^{-3}\,\mbox{S.cm}^{-1}$ at room temperature which makes this material a superionic conductor.  The corresponding mobility is inferred to be $\mbox{10}^{-3}\,\mbox{cm}^2.\mbox{S}.\mbox{V}^{-1}$. The electronic conductivity is estimated from the value of the resistance outside the temperature range where the material shows charge accumulation and is found to be lower than $\mbox{10}^{-9}\,\mbox{S.cm}^{-1}$. This extremely low electronic conductivity is consistent with band structure calculations presented in Fig.\,5 in the Supplemental material that show that the material is electronically insulating with a gap of 3.53\,eV. This proves that the reduction process due to annealing has negligible effect on the electronic doping which is at odds with what is observed in other perovskites (see e.g. \cite{Biscaras:2012} and references therein) or in memristive titanium oxide devices \cite{Strukov:2008}.

It is this combination of extremely small electronic conductivity and relatively high ionic conductivity that makes the system behave like a giant dipole.
This also implies that charge transfer at the electrode interfaces is negligible; the carriers are confined in the sample and then accumulate at the edges. 

Why is the reported colossal dielectric constant not universal to all superionic conductors? As a matter of fact, the dielectric constant of most  solid electrolytes is usually not reported to be very high. For instance sodium-beta-alumina has a dielectric constant along the c-axis, which is the non-conductive direction, of about $\sim \mbox{200}$ \cite{Pal:2009gj}. An explanation may be that, in order to protect the colossal dielectric constant,  the electronic conductivity needs to be extremely small (and the charge transfer at the interfaces immaterial). In most of the oxygen conducting electrolytes for instance, the presence of vacancies is generally known to dope the system and increase its electronic conductivity, which is detrimental to $\epsilon$ due to possible screening from the electrons or holes.  Recent results on a polymer-based solid electrolyte report a very high dielectric constant  \cite{Srivastava:2016be}, which could be explained by the above proposed mechanism, although no hysteretic behavior is reported to our knowledge. On the other hand some titanium-based perovskites have been reported to be good oxygen conductors\cite{Li:2014}. Another explanation could be that the mechanism has to imply at least two different ionic species with charges of opposite signs, as demonstrated above.

The remarkable properties  of this material make it interesting for applications, in particular for the realization of supercapacitors for energy storage \cite{Simon:2008dc, Simon:2014}.  Its high ionic conductivity and very low electronic conductivity are particularly interesting for its use as a solid electrolyte.   The above described properties were also observed on a ceramic, and work is in progress to characterize the ionic behavior in this material at the nanoscale \cite{Maier:2005}.

\textbf{Conclusion}

We have demonstrated a colossal dielectric constant (up to $10^9$, so 2 to 3 orders of magnitude larger than the current state-of-the-art in crystals) in a lamellar perovskite structure of chemical formula  \RTOd.  Evidence has been provided that this material behaves like a ferroelectric material in the temperature range $200\,\mbox{K}-320\,\mbox{K}$, although its crystal structure does not belong to pyroelectric classes. The presence of Warburg impedance elements in the permittivity at low frequency points toward the existence of ionic diffusion processes. We provide an explanation in terms of a giant dipole created by  ion migration inside the sample with the creation of a diffusion layer due to accumulation near the interfaces with the electrodes. A correlation between ionic conductivity and the polarizability character is demonstrated.  This material may therefore be qualified as a "ferro-ionet" by analogy to ferro-electrets. It realizes at the same time a colossal dielectric constant at low frequency together with a high internal ionic conduction. These findings in a crystalline solid (single crystal or ceramic) open new possibilities in particular for the realization of solid electrolytes and super capacitors.

\phantomsection


\begin{thebibliography}{10}

\bibitem{VonHippel:1946}
A.~Von~Hippel, R.G. Breckenridge, F.G. Chesley, and Tisza L.
\newblock High dielectric constant ceramics.
\newblock {\em Ind. Eng. Chem.}, 38 (1946).

\bibitem{Wul:1946}
B.~Wul and J.M. Goldman.
\newblock {\em C.R. Acad. Sci. URSS}, 51 (1946).

\bibitem{Jaffe:1971}
B.~Jaffe, W.~R. Cook, and H.~Jaffe.
\newblock {\em {Piezoelectric Ceramics}}.
\newblock Academic (1971).

\bibitem{Cohen:1998}
Ronald~E. Cohen.
\newblock {\"O}rigin of ferroeletricity in perovskite oxides.
\newblock {\em Nature}, 358:136--138 (1998).

\bibitem{Bednorz:1986}
J.G. Bednorz and K.A. Müller.
\newblock {Possible high-Tc superconductivity in the Ba-La-Cu-O system}.
\newblock {\em Z. Physik B - Condensed Matter}, 64:189--193 (1986).

\bibitem{Chu:1987}
M.~K. Wu, J.~R. Ashburn, C.~J. Torng, P.~H. Hor, R.~L. Meng, L.~Gao, Z.~J.
  Huang, Y.~Q. Wang, and C.~W. Chu.
\newblock {Superconductivity at 93 K in a new mixed-phase Y-Ba-Cu-O compound
  system at ambient pressure}.
\newblock {\em Phys. Rev. Lett.}, 58:908--910 (1987).

\bibitem{Ramirez:1997}
A.~P. {Ramirez}.
\newblock {Review Article: Colossal magnetoresistance}.
\newblock {\em Journal of Physics Condensed Matter}, 9:8171--8199 (1997).

\bibitem{ANDERSSON:1960fi}
S.~Andersson and A.~D. Wadsley.
\newblock {Five Co-ordinated Titanium in $K_2Ti_2O_5$}.
\newblock {\em Nature}, 187:499 (1960).

\bibitem{Andersson:1961vf}
S.~Andersson and A.~D. Wadsley.
\newblock {The crystal structure of $K_2Ti_2O_5$}.
\newblock {\em Acta Chem. Scand.} 15, 3 (1961).

\bibitem{VESTA}
K. Momma and F. Izumi 
\newblock {VESTA 3 for three-dimensional visualization of crystal, volumetric and morphology data}.
\newblock  {\em J. Appl. Crystallogr.}, 44, 1272-1276 (2011).

\bibitem{Warburg:1899}
Warburg E.
\newblock {Ueber die spitzenentladung}.
\newblock {\em Annu. Phys. Chem.}, 67:493 (1899).

\bibitem{Cole:1941fy}
Kenneth~S. Cole and Robert~H. Cole.
\newblock {Dispersion and Absorption in Dielectrics I. Alternating Current
  Characteristics}.
\newblock {\em The Journal of Chemical Physics}, 9(4):341 (1941).

\bibitem{Cole:1942bo}
Kenneth~S. Cole.
\newblock {Dispersion and Absorption in Dielectrics II. Direct Current
  Characteristics}.
\newblock {\em The Journal of Chemical Physics}, 10(2):98 (1942).

\bibitem{Sekiya:2000} 
T. Sekiya, K. Ichimura, M. Igarashi and S. Kurita, 
\newblock {Absorption spectra of anatase TiO$_2$ single crystals heat-treated under oxygen atmosphere}.
\newblock {\em Journal of Physics and Chemistry of Solids } \textbf{61}, 1237–1242 (2000)

\bibitem{Ramirez:2000tn}
A.~P. Ramirez, M.~A. Subramanian, M.~Gardel, G.~Blumberg, D.~Li, T.~Vogt, and
  S.~M. Shapiro.
\newblock {Giant dielectric constant response in a copper-titanate}.
\newblock {\em Solid State Communications}, 115(5):217--220 (2000).

\bibitem{Homes:2001ve}
C.~C. Homes, T.~Vogt, S.~M. Shapiro, and S.~Wakimoto.
\newblock {Optical response of high-dielectric-constant perovskite-related
  oxide}.
\newblock {\em {Science}}, 293:673 (2001).

\bibitem{Li:2005ep}
J~Li, A~W Sleight, and M~A Subramanian.
\newblock {Evidence for internal resistive barriers in a crystal of the giant
  dielectric constant material: CaCu3Ti4O12}.
\newblock {\em Solid State Communications}, 135(4):260--262 (2005).

\bibitem{Wang:2012ib}
Chun-Chang Wang, Mei-Ni Zhang, Ke-Biao Xu, and Guo-Jing Wang.
\newblock {Origin of high-temperature relaxor-like behavior in CaCu3Ti4O12}.
\newblock {\em Journal of Applied Physics}, 112(3):034109 (2012).

\bibitem{Wu:2002ko}
Junbo Wu, Ce-Wen Nan, Yuanhua Lin, and Yuan Deng.
\newblock {Giant Dielectric Permittivity Observed in Li and Ti Doped NiO}.
\newblock {\em Physical Review Letters}, 89(21):217601 (2002).

\bibitem{Larsen:1973vd}
P~K Larsen and R~Metselaar.
\newblock {Electric and Dielectric Properties of Polycrystalline Yttrium Iron
  Garnet: Space-Charge-Limited Currents in an Inhomogeneous Solid}.
\newblock {\em Physical Review B}, 8:2016 (1973).

\bibitem{Biscaras:2012}
J. Biscaras, B. Leridon, D. Colson, A. Forget and P. Monod
\newblock {Direct evidence of interchange between hole doping and Curie paramagnetism in underdoped $YBa_2Cu_3O_x$}.
\newblock {\em Physical Review B}, 85:134517 (2012).

\bibitem{Strukov:2008}
 Dmitri B. Strukov, Gregory S. Snider,  Duncan R. Stewart and R. Stanley Williams
\newblock {The missing memristor found}.
\newblock {\em Nature}, 453: 7191 (2008).

\bibitem{Pal:2009gj}
Bhola~N. Pal, Bal~Mukund Dhar, Kevin~C. See, and Howard~E. Katz.
\newblock {Solution-deposited sodium beta-alumina gate dielectrics for
  low-voltage and transparent field-effect transistors}.
\newblock {\em Nature Materials}, 8(11):898--903 (2009).

\bibitem{Srivastava:2016be}
Neelam Srivastava and Manindra Kumar.
\newblock {Ion dynamics and relaxation behavior of NaPF$_6$-doped polymer
  electrolyte systems}.
\newblock {\em Journal of Solid State Electrochemistry}, 20:1--8 (2016).

\bibitem{Li:2014}
Ming Li, Martha~J. Pietrowski, Roger A.~De Souza, Huairuo Zhang, Ian~M. Reaney,
  Stuart~N. Cook, John~A. Kilner, and Derek~C. Sinclair.
\newblock A family of oxide ion conductors based on the ferroelectric
  perovskite Na$_0.5$Bi$_0.5$TiO$_3$.
\newblock {\em Nature Materials}, 13:31 -- 35 (2014).


\bibitem{Simon:2008dc}
Patrice Simon and Yury Gogotsi.
\newblock {Materials for electrochemical capacitors}.
\newblock {\em Nature Materials}, 7(11):845--854 (2008).

\bibitem{Simon:2014}
Patrice Simon, Yury Gogotsi, and Bruce Dunn.
\newblock {Where Do Batteries End and Supercapacitors Begin?}
\newblock {\em Science}, 343(6176):1210--1211 (2014).

\bibitem{Maier:2005}
J.~Maier.
\newblock Nanoionics: ion transport and electrochemical storage in confined
  systems.
\newblock {\em Nature Materials}, 4:805 -- 815 (2005).

\end{thebibliography}

\end{document}